\documentstyle[icmp,epsfig]{article}

\def\ffrac#1#2{\textstyle{#1\over#2}\displaystyle}
\def\half {\mbox{$\textstyle {1 \over 2}$}}
\def\I{{\rm i}}
\def\e{{\rm e}}

\def\Journal#1#2#3#4{{#1} {\bf #2}, #3 (#4)}

\def\NPB{{\em Nucl. Phys.} B}

\def\JPA{{\em J. Phys.} A}
\def\PRL{\em Phys. Rev. Lett.}
\def\NPB{{\em Nucl. Phys.} B}
\def\JSP{\em J. Stat. Phys.}
\def\EPL{\em Europhys. Lett.}
\def\TMP{\em Theor. Math. Phys.}

\def\be{\begin{equation}}
\def\ee{\end{equation}}
\def\bea{\begin{eqnarray}}
\def\eea{\end{eqnarray}}

\begin{document}

\title{THE INTEGRABLE OPEN $XXZ$ CHAIN\\ WITH BROKEN $Z_2$ SYMMETRY} 

\author{M. T. BATCHELOR}

\address{Department of Mathematics, School of Mathematical Sciences,\\
The Australian National University, Canberra ACT 0200, Australia\\
E-mail: murrayb@maths.anu.edu.au} 


\maketitle\abstracts{
The hamiltonian of an asymmetric diffusion process with
injection and ejection of particles at the ends of a
chain of finite length is known to be relevant to
that of the spin-$\ffrac{1}{2}$ XXZ chain with integrable 
boundary terms. However, the inclusion of boundary 
sources and sinks of particles breaks 
the arrow-reversal symmetry necessary for solution via 
the usual Bethe Ansatz approach. Developments in solving 
the model in the absence of arrow-reversal symmetry are discussed.}

\section{Introduction}

The spin-$\ffrac{1}{2}$ $XXZ$ Heisenberg chain is the canonical 
example of an integrable quantum hamiltonian. 
The integrability is assured by the triangle or factorisation equations,
\be
R_{12}(u-v) R_{13}(u) R_{23}(v) = R_{23}(v) R_{13}(u) R_{12}(u-v).
\ee
Among other interpretations, this relation is often described 
pictorially in terms of the scattering of three particles.  
This notion of integrability has been extended to include the 
presence of a boundary,\,\cite{C,Sk,MN} with a similar interpretation
in terms of particles reflecting from a wall.
The reflection equation is
\be
R_{12}(u-v) K_1(u) R_{21}(u+v) K_2(v) = K_2(v) R_{12}(u+v) K_1(u) R_{21}(u-v).
\ee
Our interest here lies in the fact that the above $R$ and $K$ matrices
define the Boltzmann weights for exactly solvable two-dimensional lattice 
models.
In particular, the reflection equation follows as the condition for two
Sklyanin double-row transfer matrices (Fig. 1(a)) to commute.
The boundary vertex weights can be simply written in terms of the $K$-matrix 
elements.\,\cite{YB}

Consider the concrete example of the six-vertex model and the related
$XXZ$ spin chain, for which the $R$ and $K$ matrices are given by
\be
R=\left( \begin{array}{cccc}   ~a~ & ~ & ~ & ~      \\
                               ~ & \sin u~  & ~\sin \lambda  & ~  \\
                               ~ & \sin \lambda~ & ~\sin u & ~  \\
                               ~ & ~ & ~ & ~a~ 
\end{array} \right), \quad K = \left( \begin{array}{cc}
      k \sin(\xi+u)~ & ~\mu \sin(2u) \\
      \nu \sin(2u)~& ~k \sin(\xi-u)
\end{array} \right)
\ee
where $a=\sin(\lambda-u)$ with $k, \xi, \mu, \nu$ free parameters.
This $K$-matrix was found only quite recently.\,\cite{Sk,dV}

\begin{figure}
\centerline{
\epsfxsize=2.5in
\epsfbox{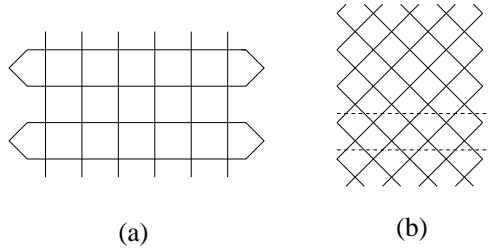}}
\caption{(a) Sklyanin double-row transfer matrices and (b) 
the open diagonal lattice.}
\end{figure}

The vertex weights along the right boundary in Fig. 1(b) follow as 
\be
\begin{picture}(200,55)
\put( 20,30){\line( 1, 1){ 10}}
\put( 20,30){\vector( 1, 1){7}}
\put( 30,40){\line(-1, 1){ 10}}
\put( 30,40){\vector(-1, 1){7}}
\put( 35,38){$= \,k \sin(\xi+\half u)$,}
\put(120,30){\line( 1, 1){ 10}}
\put(120,30){\vector( 1, 1){7}}
\put(130,40){\line(-1, 1){ 10}}
\put(120,50){\vector(1,-1){7}}
\put(135,38){$= \,\mu \sin(u)$,}
\put( 20, 0){\line( 1, 1){ 10}}
\put( 30,10){\vector(-1,-1){7}}
\put( 30,10){\line(-1, 1){ 10}}
\put( 20,20){\vector(1,-1){7}}
\put( 35, 8){$= \,k \sin(\xi-\half u)$,}
\put(120, 0){\line( 1, 1){ 10}}
\put(130,10){\vector(-1,-1){7}}
\put(130,10){\line(-1, 1){ 10}}
\put(130,10){\vector(-1, 1){7}}
\put(135, 8){$= \,\nu \sin(u)$.}
\end{picture}
\label{bws}
\ee
In general there is a set of parameters $k_\pm, \xi_\pm, \mu_\pm, \nu_\pm$
for each boundary.
These correspond to the two free ends of the related spin chain, with 
hamiltonian\,\cite{dV}
\bea
\lefteqn{
H = \sum_{j=1}^{N-1} \left( \sigma_j^x \sigma_{j+1}^x + 
\sigma_j^y \sigma_{j+1}^y + \Delta \,\, 
\sigma_j^z \sigma_{j+1}^z \right)} \nonumber\\
& & + \left(p_- \sigma_1^z + p_+ \sigma_N^z 
+ c_- \sigma_1^- + c_+ \sigma_N^- + d_- \sigma_1^+ + d_+ \sigma_N^+ \right), 
\label{gen}
\eea
where ${\boldmath \sigma} =(\sigma^x,\sigma^y,\sigma^z)$ are the Pauli
matrices, with $\sigma^\pm = \ffrac{1}{2}(\sigma^x \pm \I\, \sigma^y)$.
The parameters $p_\pm$ (related to $\xi_\pm$) control the strength
of the $z$-component of the surface fields. The parameters 
$c_\pm$ and $d_\pm$ (related to $\nu_\pm$ and $\mu_\pm$) are the terms 
responsible for breaking the familiar ``up-down'' $Z_2$ symmetry.

Despite some intensive effort, the six-vertex model with boundary weights 
defined by the above $K$-matrix, and equivalently the $XXZ$ chain with the 
above boundary terms, still defy an exact solution, by which I mean that
both the transfer matrix and the hamiltonian remain to be diagonalised.
The major obstacle is readily apparent -- the boundary terms arising
from the non-diagonal elements of the $K$-matrix break  
the $Z_2$ symmetry by the introduction
of sources and sinks of arrows or particles. The total spin along
a row of vertical bonds in the vertex model is no longer a conserved
quantity.
Such a good quantum number is essential input into the Bethe Ansatz
method of solution.
%

\section{Special cases}

\subsection{Solution for $\mu=\nu=0$.}

Consider first the spin chain, which has been solved for the corresponding 
choice of $c_\pm=d_\pm=0$,\,\cite{Sk,ABBBQ}
\be
H_1(\Delta,p_-,p_+) = \sum_{j=1}^{N-1} \left( \sigma_j^x \sigma_{j+1}^x +
\sigma_j^y \sigma_{j+1}^y + \Delta \,\, \sigma_j^z \sigma_{j+1}^z \right) 
+ p_- \sigma_1^z + p_+ \sigma_N^z
\ee
The eigenvalues are given by
\be
E = (N-1) \Delta + 4 \sum_{j=1}^n (\cos k_j - \Delta)
\ee
with the Bethe equations
\be
\e^{\I\, 2(L-1) k_j} = \frac{f_{-k_j}(p_-,\Delta) f_{-k_j}(p_+,\Delta)}
                          {f_{k_j}(p_-,\Delta) f_{k_j}(p_+,\Delta)}
{\prod_{l=1}^{n}}' \frac{S(k_l,k_j)}{S(k_l,-k_j)}
\label{BE}
\ee
where $f_k(a,b) = a - b + \e^{\I k}$ and the prime denotes $\l \ne j$.
Also $S(p,q) = s(p,q)/s(q,p)$, where
$s(p,q) = 1 - 2 \Delta \e^{\I q} + \e^{\I(p+q)}$.
This hamiltonian has $U_q[su(2)]$-symmetry\,\cite{PS} when 
$\Delta=-\cos \lambda$ with $p_- = -p_+ = \I \sin \lambda$,
known also as the `Potts case'.\,\cite{ABBBQ}

The six-vertex model on the particular open lattice shown in Fig. 1(b) 
has been considered by a number of authors. For this lattice there is a 
diagonal-to-diagonal transfer matrix as indicated.
It was solved by means of the
co-ordinate Bethe Ansatz for boundary weights corresponding to 
$\mu=\nu=0$.\,\cite{OB}
The Sklyanin double-row transfer matrix (Fig. 1(a)) was also 
diagonalised for $\mu=\nu=0$ via the algebraic Bethe Ansatz.\,\cite{Sk}
It is worth noting that a similar double-row transfer matrix was 
diagonalised earlier by Baxter,\,\cite{Bax} who used the Bethe Ansatz 
solution to obtain the surface free energy.
Later it was shown how to pass from the double-row transfer matrix  
to the diagonal-to-diagonal transfer matrix by means of a
special choice of alternating inhomogeneities.\,\cite{YB,DV}
The more general double-row transfer matrix was also diagonalised.
A key point is that the diagonal-to-diagonal transfer matrix does not
commute -- rather the underlying integrability lies in the
commutation of the Sklyanin transfer matrix with 
alternating inhomogeneities.
A number of other models have also been solved on the lattice in Fig. 1(b)
with diagonal $K$-matrices.\,\cite{YB}
Solutions of the various quantum invariant spin chains are also known.

I shall not reproduce the solution of the six-vertex model with
diagonal $K$-matrices here. However, it is worth noting that the
parametrisation of the diagonal boundary weights in (\ref{bws}),
given via the solution of the reflection equation, is particularly
convenient when substituted into the corresponding weights $d$ and $e$
of Owczarek and Baxter\,\cite{OB}.    

\subsection{The case $\mu=0$ or $\nu=0$.}

Given the impasse on solving the general problem it came as quite a 
surprise when a solution was reported for the hamiltonian\,\cite{SSa,SSb} 
\be
H_2(\alpha,\beta,\gamma,\delta) = 
\sum_{j=1}^{N-1} \left( \sigma_j^x \sigma_{j+1}^x +
\sigma_j^y \sigma_{j+1}^y + \sigma_j^z \sigma_{j+1}^z \right) + H_s
\label{ss}
\ee
The surface term $H_s$ following from the stochastic dynamics of symmetric
hopping of particles in one dimension, with particle injection and ejection
at the boundaries, is
\bea
H_s &=& (\alpha+\gamma)(\sigma_1^x-1) + (\gamma-\alpha)
(\I\,\sigma_1^y-\sigma_1^z)
+ \nonumber\\
&& \quad (\beta+\delta)(\sigma_L^x-1) + (\beta-\delta)
(\I\,\sigma_L^y-\sigma_L^z)
\nonumber\\
&=& 2 \alpha\, \sigma_1^- + 2 \gamma\, \sigma_1^+ - (\alpha+\gamma) -
(\gamma-\alpha) \sigma_1^z +
\nonumber\\
&& \quad 2 \beta\, \sigma_L^+ + 2 \delta\, \sigma_L^- 
- (\beta+\delta) - (\beta-\delta) \sigma_L^z
\eea
This is clearly a special case of the general hamiltonian (\ref{gen}).
Bethe equations were obtained\,\cite{SSb} similar in form to those 
given in (\ref{BE}) with $\Delta=1$.
However, checking the operator algebra in Ref.\,\cite{SSb} reveals that the
solution is precisely that given in (\ref{BE}), with the identification
$\Delta=1$, $p_-=\alpha+\gamma$, $p_+=\beta+\delta$.     
Moreover, numerical diagonalisation reveals the equivalence\,\cite{AB}
\be
H_1(1,\alpha+\gamma,\beta+\delta) = H_2(\alpha,\beta,\gamma,\delta)
\ee
This is a consequence of the eigenvalues of 
\be
H_1(\Delta,p_-,p_+) + d_- \sigma_1^+ + d_+ \sigma_L^+ 
\ee
being independent of the variables $d_\pm$ (and similarly for
$\sigma_1^-$ and $\sigma_L^-$).\,\cite{AB}

Similar considerations apply to the six-vertex model. Certain 
combinations of sources and sinks at the boundary do not change
the eigenspectrum. In particular, a boundary sink of arbitrary
weight may be included on both edges, or equivalently a boundary source
on both edges. Such behaviour is consistent with inversion 
relations\,\cite{Z} in which the off-diagonal terms involve the 
prefactors $\mu_- \nu_-$ and $\mu_+ \nu_+$.

\section{Concluding remarks}

Obtaining the solution with the general $K$-matrix remains an 
outstanding problem. It may still be that the solution can be obtained 
via the pair propagation through a vertex technique used for the
six-vertex model with antiperiodic boundary conditions, where the
$Z_2$ symmetry is also broken.\,\cite{BBOY}

The implications of the special solutions with sources and sinks at
the boundary to the diffusion problems remain to be fully explored.
These include the isotropic hamiltonian (\ref{ss}) of relevance
to symmetric hopping with particle injection and ejection at the
boundary\,\cite{SSa,SSb} (see also the vertex model in Ref.\,\cite{HP}). 
It is clear that the corresponding anisotropic $XXZ$ chain
enjoys a similar property, however its precise meaning in terms
of the asymmetric hopping of particles needs to be clarified. 

\section*{Acknowledgments}
It is a pleasure to thank F.C. Alcaraz, R.J. Baxter and Y.K. Zhou 
for helpful discussions and correspondence. 
I also thank the Australian Research Council for support.

\end{document}